\documentclass[aps,showpacs]{revtex4}
\usepackage{epsfig,amsmath,amsfonts,wasysym}
\usepackage{dsfont}
\newcommand{\fr}[2]{{\frac{#1}{ #2}}}
\newcommand{\Ref}[1]{(\ref{#1})}
\newcommand{\be}{\begin{equation}}
\newcommand{\ee}{\end{equation}}
\newcommand{\bn}{\begin{eqnarray}}
\newcommand{\en}{\end{eqnarray}}
\newcommand{\bd}{\begin{displaymath} }
\newcommand{\ed}{\end{displaymath}}
\newcommand{\bnn}{\begin{eqnarray*}}
\newcommand{\enn}{\end{eqnarray*}}
\newcommand{\bs}{\begin{subequations}}
\newcommand{\es}{\end{subequations}}
\newcommand{\adb}{\allowdisplaybreaks }

\begin{document}

\title{Self-force on a scalar particle in a class of wormhole spacetimes}
\author{V.B. Bezerra$^{a}$\footnote{e-mail: valdir@fisica.ufpb.br} and \
Nail R. Khusnutdinov$^{a,b,c}$\footnote{e-mail: nail@fisica.ufpb.br}}

\address{a) Departamento de F\'{\i}sica, Universidade Federal da
Para\'{\i}ba, Caixa Postal 5008, CEP 58051-970 Jo\~ao Pessoa, Pb,
Brazil\\
b) Kazan State University, Kremlevskaya 18, Kazan, 420008, Russia,\\
c) Tatar State University of Humanities and Education, Tatarstan 2,
Kazan 420021, Russia }

\begin{abstract}
We consider the self-energy and the self-force for scalar massive and
massless particles at rest in the wormhole space-time. We develop a general
approach to obtain the self-force and apply it to the two specific profiles of
the wormhole throat, namely, with singular and with smooth curvature. We found
that the self-force changes its sign at the point where nonminimal coupling $\xi
= 1/8$ (for massless case) and it tends to infinity for specific values of
$\xi$.  It may be attractive as well repulsive depending on the profile of the
throat. For massless particle and minimal coupling case the electromagnetic
results are recovered. 
\end{abstract}


\maketitle

\section{Introduction}
The wormholes are topological bridges which connect different universes or
different parts of the same universe. These kind of tunnels in space-time have
been appeared in different contexts of physics: in the analysis of black
hole backgrounds \cite{Flamm:1916:BzEG,Einstein:1935:PPGTR}, as an idea to
construct "charge without charge"\/ and "mass without mass"\/
\cite{Wheeler:1955:G,Wheeler:1960:Ngg}, and as a possibility for time machine
\cite{Morris:1988:Wstuitttgr,Morris:1988:WTMWEC}. The wormholes require an
amount of exotic matter which breaks energy conditions of a solution of
Einstein equations. There are many exotic ideas how to produce
exotic matter and how much exotic matter one needs to make possible existence of
wormhole. Carefull discussion of wormhole's geometry and physics may be found in
the Visser book \cite{Visser:1995:LWfEtH}, in the review by Lobo
\cite{Lobo:2007:EsGRTwwds}, as well as in paper \cite{Kardashev:2007:AW}
related with astrophysical implementation of wormholes. Among interesting
publications we can mention Ref. \cite{Doroshkevich:2008:Prtw} which analysed
numerically  the process of the passage of a radiation pulse through a wormhole
and the subsequent evolution of the wormhole. It was shown that the wormhole is
unstable and it is transformed into a spacetime with horizon. The analysis was
made for normal as well as exotic matter pulse formed by scalar fields. 

It is well known that a particle in curved spacetimes may interact with
the gravitational background by specific interaction due to self-force
\cite{DeWitt:1960:Rdgf}. The origin of this self-force is associated
with nonlocal structure of the field, the source of which the particle is.
The self-force may be the unique gravitational interaction on a particle as it
happens for particles in the space-time of a cosmic string
\cite{Linet:1986:FCSCS}. In this case there is no gravitational interaction
particle with cosmic string, but nevertheless there exist the repulsive
self-interaction force. In contrast to standard self-interaction Dirac-Lorentz
force  \cite{Dirac:1938:Ctre}, the self-interaction force in
curved space-times depends on all history of the particle and it is usually
non-zero even for a particle at rest. For a particle at rest it may be
found as coincidence limit of the renormalized Green function
\cite{Smith:1989:GEISCSCQFSVF}. A detailed discussion of the self-force maybe
found in reviews \cite{Poisson:2004:mppcs,Khusnutdinov:2005:Psegf}.

In a recent paper \cite{Khusnutdinov:2007:Scpwst} the self-force for an 
electromagnetically charged particle at rest in the static wormhole
background was analyzed in detail. The general expression for self-energy for
arbitrary profile of the throat was obtained. It was shown that the particle is
attracted by the wormhole and this effect may has astrophysical
applications. For a specific profile of the throat the result was confirmed by
Linet in Ref.\cite{Linet:2007:Ewg} using a different approach. There is also
another approach for this question which was developed by Krasnikov in recent
publication \cite{Krasnikov:2008:Eipcw} for a specific profile of the throat.
The difference of results connects with understanding the self-force itself. The
self-force for scalar particle reveals peculiarities
\cite{Zelnikov:1982:Egpecp,%
Burko:2000:SfscSs,Wiseman:2000:ssstoSbh,Pfenning:2002:Segswcs} due to
nonminimal coupling of the scalar and the gravitational fields. For example,
in Schwarzschild spacetime the self-force on a particle at rest is zero for
minimal coupling \cite{Zelnikov:1982:Egpecp}. In the present paper we
analyse in detail the self-force on a scalar particle at rest in the background
of wormholes. We found that the self-force has crucial dependence on
the nonminimal coupling constant $\xi$: it is zero for $\xi = 1/8$ in massless
case and it tends to infinity for specific values  of $\xi$. 

The organization of this paper is as follows. In Sec. II we develop our approach
and consider the origin for the divergence of self-energy for specific values of
$\xi$, from the point of view quantum mechanics. In Sec. III we consider
the self-energy and self-force for massive and massless cases for two
specific profiles and for a general profile of the throat. Section IV is devoted to
the discussion of the results. Throughout this paper we use units $c=G=1$.

\section{Approach}\label{Sec:2}
Firstly let us say some words about the background under consideration. We
use the following line element of the spherically symmetric wormhole space-time
\begin{equation}
 ds^2 = -dt^2 + d\rho^2 + r^2(\rho) d\Omega^2,\label{ds^2}
\end{equation}
where the profile function $r(\rho)$ describes the shape of the throat and
$d\Omega^2 = d\theta^2 + \sin^2\theta d\varphi^2$. The
variables are defined in the following ranges: $t,\rho \in \mathds{R}$,
$\theta \in [0,\pi]$ and $\varphi \in [0,2\pi]$. The radius of the throat is
defined by $a =r(0)$ and $r'(0)=0$. The nonzero components
of the Ricci tensor and scalar curvature read:
\begin{eqnarray}
 R^\rho_\rho &=& -\frac{2r''}{r},\adb\nonumber\\
 R^\theta_\theta = R^\varphi_\varphi &=& -\frac{-1+r'^2 + rr''}{r^2}\adb\\
 R &=& -\frac{2(-1+r'^2 + 2rr'')}{r^2}.\nonumber
\end{eqnarray}
The three dimensional section corresponding to constant time of this space-time
is conformally flat. Indeed, let us consider the $3D$ flat space in spherical
coordinates $\tilde r,\theta,\varphi$. Thus, we can write
\begin{equation}
 dl^2_{fl} = d\tilde r^2 + \tilde r^2 d\Omega^2.
\end{equation}
Let us choose a new radial coordinate $\rho$ by the relation $\tilde r =
r(\rho)e^{\sigma (\rho)}$ with 
\begin{equation}
 \sigma = \pm \int^\rho \frac{dx }{r(x)} - \ln r(\rho).
\end{equation}
Using this  coordinate system we obtain 
\begin{equation}
 dl^2_{fl} = e^{2\sigma}(d\rho^2 + r(\rho)^2 d\Omega^2) = e^{2\sigma}dl^2_{wh}.
\label{conformal}
\end{equation}
Therefore $g_{ik}^{wh} = e^{-2\sigma}g_{ik}^{fl}$ and the $3D$ section is
conformally flat. For this reason it is expected \cite{Hobbs:1968:Rdcfu} that
the self-force is zero for $\xi = 1/8$ and $m=0$ as will
be shown by manifest calculations. More information about the wormhole's
space-time may be found in book \cite{Visser:1995:LWfEtH} and review
\cite{Lobo:2007:EsGRTwwds}. Some figures for specific profile of
throat are in Ref. \cite{Khusnutdinov:2003:Sw}, and the physics in this spacetime is
discussed in Ref. \cite{Doroshkevich:2008:prtwAR}.

Let us consider a massive scalar field, $\phi$, with scalar source, $j$, which
lives in the wormhole background with non-conformal coupling $\xi$. The
action consists of two part, the first one is for the field itself and the
second one describes the interaction of the source, a scalar charge $e$, with
the field and is given by
\begin{equation}
 S = -\frac{1}{8\pi} \int (\phi_{,\mu}\phi^{,\mu} + \xi R \phi^2 + m^2 \phi^2)
\sqrt{-g} d^4x + \int j\phi \sqrt{-g}d^4x.
 \end{equation}
The variation of this action with respect to the metric gives the energy momentum
tensor of the field with contribution due to the interaction field with charge
\begin{equation}
 T_{\mu\nu} = j\phi g_{\mu\nu} + \frac{1}{4\pi}
\left(\phi_{;\mu} \phi_{;\nu} - \frac{1}{2}g_{\mu\nu}
\phi_{;\sigma}\phi^{;\sigma} - \frac{1}{2}m^2 g_{\mu\nu}\phi^2 \right) +
\frac{\xi}{4\pi}\left(G_{\mu\nu} \phi^2 + g_{\mu\nu} \square \phi^2 - \phi^2_{\
;\mu\nu}\right),
\end{equation}
while the variation with respect to the field gives the equation of motion 
\begin{equation}
 (\square -\xi R - m^2)\phi = -4\pi j,
\end{equation}
with scalar current 
\begin{equation}
   j(x) =  e \int \delta^{(4)}(x-x_p(\tau)) \frac{d\tau}{\sqrt{-g}}.
\end{equation}

We consider only the case in which the particle is at rest at the point $x_p$ in
the wormhole space-time. This means that there is no dependence on the time and
the equation of motion for the field has the following form
\begin{equation}
 (\triangle -m^2 -\xi R)\phi(x) = -4\pi j = -\frac{4\pi e}{\sqrt{-g}}
\delta^{(3)}(x-x_p).\label{eq:phi}
\end{equation}
From the general point of view the full energy of particle reads 
\begin{equation}
 E = -\int T_{\mu\nu} \xi^\mu d\Sigma^\nu,
\end{equation}
where $\Sigma$ is $3$-surface with $\xi^\mu$ as a normal. The spacetime under
consideration possesses the time-like Killing vector $\xi^\mu = \delta^\mu_0$.
Thus choosing the hypersurface of constant time we
obtain 
\begin{equation}
E= -\int T_{00} \sqrt{-g}d^3 x,
\end{equation}
where for the static case under consideration we have 
\begin{equation}
 T_{00} = -j\phi + \frac{1}{8\pi} \left(\phi_{;i}\phi^{;i} + (\xi R + m^2)
\phi^2 \right) - \frac{\xi}{2\pi}\left(\phi_{;i}\phi^{;i} + \phi \triangle
\phi\right).
\end{equation}

Integrating by part and taking into account the equation of motion
(\ref{eq:phi}) we get
\begin{eqnarray*}
E &=& \frac 12 \int j\phi \sqrt{-g} d^3x = \frac 12 \int\int j(x) G(x,x')
j(x') \sqrt{-g(x)}\sqrt{-g(x')} d^3x d^3 x'\adb\nonumber\\ 
&=& \frac{e^2}{2}G(x_p,x_p).
\end{eqnarray*}
The Green's function $G(x,x')$ which we need for calculation self-force is 
divergent in the coincidence limit, ${x'} \to {x}$. Several
methods to obtain a finite result are known. The most simple way is to consider
total mass as a sum of observed finite mass and an infinite electromagnetic
contribution. Usually this procedure is called "classical renormalization"\
because there is no Planck constant in the divergent term. Dirac
\cite{Dirac:1938:Ctre} suggested to consider radiative Green's function to
calculate the self-force. Since the radiative Green's function is the difference
between retarded and advanced Green's functions, singular contribution cancels
out and we obtain a finite result. There is also an axiomatic approach suggested
by Quinn and Wald \cite{Quinn:1997:Aategrrpcs}. In the framework of this
approach one obtains finite expression by using a "comparison"\ axiom. This
approach was used in Refs. \cite{Burko:2000:?,Burko:2001:?} for a
specific space-time. 

There is a problem with renormalization for massive uncharged particle (see
Ref. \cite{Mino:1997:Grrtpm}). The self-force for charged particle is obtained
by classical renormalization of particle's mass, $m\to m + A e^2$, where $A$
is some infinite constant. Therefore for massive and uncharged particle we have
to renormalize mass by using the same mass of particle. To solve this problem
in the Ref. \cite{Mino:1997:Grrtpm} has suggested to consider particle as
Schwarzschild solution in external gravitational field. In this way it is
possible to obtain finite result in lower power of mass. More rigorous
derivation has been done in Ref. \cite{Gralla:2008:RDGS}. 

We will use a general approach to renormalization in curved space-time
\cite{Birrell:1982:QFCS}, which means subtraction of the first terms from
DeWitt-Schwinger asymptotic expansion of the Green's function. In general there
are two kinds of divergences in this expansion, namely, pole and logarithmic
ones \cite{Christensen:1978:Rrcgps}. In three-dimensional case which we are
interested in there is only pole divergence, while the logarithmic term is
absent. The singular part of the Green's function, which must be subtracted, has the
following form (in $3D$ case)
\begin{displaymath}
 G^{sing} = \frac{1}{4\pi} \left\{\frac{\triangle^{1/2}}{\sqrt{2\sigma}} +
m\right\},
\end{displaymath}
where $\sigma$ is half of the square of geodesic distance and
$\triangle$ is the DeWitt-Morette determinant. If we take the coincidence
limit for angular variables then these quantities are easily
calculated using the metric (\ref{ds^2}): $\sigma = (\rho -
\rho')^2/2$ and $\triangle = 1$. Thus to carry out renormalization
we should subtract the singular part of the Green's function,
which has the following form:
\begin{displaymath}
G^{sing} = \frac{1}{4\pi} \frac{1}{|\rho - \rho'|} + \frac{m}{4\pi}.
\end{displaymath}
This approach was used many times in different curved backgrounds
(e.g. see \cite{Khusnutdinov:2005:Psegf}). Therefore the self-energy $U$
for particle at rest at the point $x_p$ has the following form  
\begin{eqnarray}
U(x_p) &=& \frac{e^2}{2}G^{reg}(x_p,x_p) =
\frac{e^2}{2}\lim_{x'\to x_p}\left\{G(x_p,x')-G^{sing}(x_p,x')
\right\}.\label{Ureg}
\end{eqnarray}
For simplification of notations we will denote hereinafter the position of the
particle by $x=x_p$.

In order to find the self-energy we have to calculate
the $3D$ Green function which obeys the equation 
\begin{equation}
 (\triangle -m^2 -\xi R)G(\mathbf{x};\mathbf{x'}) =
-\frac{\delta^{(3)}(x-x')}{\sqrt{-g}},
\end{equation}
and then adopt the renormalization procedure. Due to spherical symmetry we
can expand the Green function using spherical functions as 
\begin{equation}
 G(\mathbf{x};\mathbf{x'}) = \sum_{l=0}^\infty \sum_{m=-l}^{+l}
Y^*_{l,m}(\Omega) Y_{l,m}(\Omega) g_l(\rho,\rho') = \sum_{l=0}^\infty
\frac{2l+1}{4\pi} P_l(\cos\gamma)g_l(\rho,\rho').\label{GreenDefinition}
\end{equation}
Here $P_l$ are the Legendre polynomials and $\cos\gamma = \cos\theta
\cos\theta' + \sin\theta\sin\theta' \cos(\varphi - \varphi')$.

The radial Green function,$g_l(\rho,\rho')$, obeys the following equation
\begin{equation}
g_l'' + \frac{2r'}{r}g_l' - \left(m^2 + \frac{l(l+1)}{r^2} + \xi
R\right)g_l = -\frac{\delta(\rho-\rho')}{r^2}.
\end{equation}

Differently from the electromagnetic case \cite{Khusnutdinov:2007:Scpwst} we
have a contribution which arises from the non-minimal coupling, even for
massless case. 

Now, we adopt the approach developed in Ref. \cite{Khusnutdinov:2007:Scpwst}
with some modifications. We represent the Green function in the following form 
\begin{equation}
 g_l = \theta(\rho-\rho') \Psi_2(\rho) \Psi_1(\rho') + \theta(\rho'-\rho)
\Psi_2(\rho') \Psi_1(\rho),
\end{equation}
where $\Psi_1$ and $\Psi_2$ are independent solutions of the corresponding
homogeneous
equation
\begin{equation}
 \Psi'' + \frac{2r'}{r}\Psi' - \left(m^2 + \frac{l(l+1)}{r^2} + \xi
R\right)\Psi = 0,\label{Psi}
\end{equation}
with boundary conditions
\begin{equation}
 \lim_{\rho\to+\infty} \Psi_2 = 0,\ \lim_{\rho\to+\infty} \Psi_1 \not = 0,
\label{boundarypsi}
\end{equation}
and Wronskian condition
\begin{equation}
 \Psi_1\Psi_2' - \Psi_2\Psi_1' = -\frac{1}{r^2}.
\end{equation}

It is worthily calling attention to the fact that if we change the function
$\Psi = \Phi/r$ in Eq. (\ref{Psi}), we obtain 
\begin{equation}
 \Phi'' + \left(-m^2 - \frac{l(l+1)}{r^2} - \xi
R - \frac{r''}{r}\right)\Phi = 0.\label{eq:Phi}
\end{equation}
From the quantum mechanical point of view, Eq. (\ref{eq:Phi}) describes a
quantum particle in the potential
\begin{equation}
 V = \xi R + \frac{r''}{r} 
\end{equation}
Let us consider this potential far from the wormhole's throat, $\rho \gg a$. The
behavior of the potential crucially depends on the profile function $r(\rho)$.
We divide the profiles of the throat in two large classes depending on its behavior at
infinity. The first class may be called as wormhole without parameter of the
throat's length. In this case the expansion of the profile function over
$\rho\to\infty$ has a polynomial form 
\begin{equation}
 r = \rho + \sum_{k=n}^\infty b_k \rho^{-k},
\end{equation}
which starts from $\rho^{-n}$ with some integer $n\geq 0$. The potential far
from the wormhole's throat has
the following expansion
\begin{equation}
 \xi R + \frac{r''}{r} = \frac{4n(\xi_c - \xi)b_n}{\rho^{n+3}} + \cdots 
\end{equation}
where $\xi_c = (n+1)/4n$. The point $\xi_c$ is crucial because at this
point the potential changes its sign. The smaller $n$, the greater the critical
value of $\xi$ and vice-verse. The case $n = 0$ in some sense corresponds to
the following "weak"\/ dependence of the profile of the throat on $\rho$
at infinity 
\begin{equation}
 r(\rho) = \rho +a + \frac{b}{(\ln {\rho})^n}
\end{equation}
with $n>0$. Indeed, in this case there is no critical value of $\xi$. This
kind of expansion corresponds to $n\to0$ in the above case and $\xi_c \to
\infty$.

Another kind of wormholes have a dimensional parameter, $\tau$, which describes
the throat's length. In this case the expansion for $\rho\to\infty$ has
the following form 
\begin{equation}
 r(\rho) = \rho + a + c_n \rho^n e^{-\frac{\rho}{\tau}},\ n\geq 0,
\end{equation}
and the space-time becomes flat exponentially. The expansion of
the potential starts from the following term
\begin{eqnarray}
 \xi R + \frac{r''}{r} &=& c_n\frac{1-4\xi}{\tau^2}\rho^{n-1}e^{-\frac
\rho\tau} 
+ \cdots .
\end{eqnarray}
Therefore the critical value of $\xi$ is $1/4$ and it does not depend
on $n$. Therefore, we claim the following statement: The critical value for
the first kind of throat is $\xi_c = (n+1)/4n$ if
$r-\rho \sim \rho^{-n}$, and for second kind is $\xi_c = 1/4$. Obviously the
delta function is a short range potential and it belongs to the second kind of
wormhole as we will see later. In the mixed case the main role is played by the
polynomial part of the expansion. 

Let us consider specific examples. For $r=\sqrt{\rho^2 + a^2}$ we have the first
kind of throat and $n=1,\ b_1 = 1/2$, the critical value of $\xi_c$ is $1/2$.
For the profiles 
\begin{eqnarray*}
 r &=& \rho \coth \frac{\rho}{\tau} + a - \tau,\adb\\
 r &=& \rho \tanh \frac{\rho}{\tau} + a,
\end{eqnarray*}
we have correspondingly $n=1,b=1/2$ and $n=1,b=-1/2$ and therefore
the critical value $\xi_c = 1/4$. The case of a singular potential is a 
limiting case of shortest length of the throat and it belongs to the second case.
Therefore, we can expect peculiarities for $\xi \approx
\xi_c$. Below we will see them in manifest forms.  

Let us consider the case of singular scalar
curvature separately. The point is that for the throat profile $r = |\rho| + a$,
the scalar curvature reads
\begin{equation}
 R = -\frac{8}{a}\delta (\rho)
\end{equation}
and we have the following radial equation 
\begin{equation}
 \Psi'' + \frac{2r'}{r}\Psi' - \left(m^2 + \frac{l(l+1)}{r^2} -\frac{8
\xi}{a}\delta (\rho)\right)\Psi = 0.\label{PsiSing}
\end{equation}
Integrating this equation around $\rho = 0$ we obtain the following matching
conditions at the throat
\begin{eqnarray}
 \Psi(+0) - \Psi(-0) &=& 0,\adb\nonumber\\
 \Psi'(+0) - \Psi'(-0) &=& -\frac{8\xi}{a} \Psi(+0).\label{boundarycond}
\end{eqnarray}

Let us now represent the solutions $\Psi$ as linear combinations of two
independent solutions in each domain of the wormhole spacetime, "+"\/ and "--",
which correspond to the signs of $\rho$. In each domain we find two independent
solutions $\phi^1_{\pm}$ and $\phi^2_{\pm}$ with the condition that $\phi^2_+$
falls down for $\rho\to\infty$ and the Wronskian condition 
\begin{equation}
 W(\phi^1_\pm,\phi^2_\pm) = \frac{A_\pm}{r^2},
\end{equation}
where $A_\pm$ are constants. Therefore, we have in general
\begin{eqnarray*}
\Psi_1 &=&\left\{ \begin{array}{lc}
                  \alpha^1_+ \phi^1_+ + \beta^1_+ \phi^2_+, & \rho >0 \\
                  \alpha^1_- \phi^1_- + \beta^1_- \phi^2_-, &
                  \rho <0
                \end{array} \right., \adb \\
\Psi_2 &=&\left\{ \begin{array}{lc}
                  \alpha^2_+ \phi^1_+ + \beta^2_+ \phi^2_+, & \rho >0 \\
                  \alpha^2_- \phi^1_- + \beta^2_- \phi^2_-, &
                  \rho <0
                \end{array} \right. .
\end{eqnarray*}
The Wronskian condition implies the following constraints on the coefficients:
\begin{displaymath}
\alpha^1_\pm \beta^2_\pm -\beta^1_\pm \alpha^2_\pm = -\frac 1{A_\pm}.
\end{displaymath}

Taking into account the boundary conditions (\ref{boundarycond}), then the
solutions $\Psi_1$ and $\Psi_2$ turns into the following forms
\begin{eqnarray*}
\Psi_1 &=& \alpha^1_- \tilde{\phi}^1 + \beta^1_- \tilde{\phi}^2, \adb \\
\Psi_2 &=&  \alpha^2_+ \hat{\phi}^1 + \beta^2_+ \hat{\phi}^2,
\end{eqnarray*}
where
\begin{eqnarray*}
\tilde{\phi}^1 &=& \left\{\begin{array}{cl}
\phi^1_+\left\{\fr{W(\phi_-^1,\phi_+^2)}{W(\phi_+^1,\phi_+^2)} +
\frac{8\xi\phi_-^1\phi_+^2 }{aW(\phi_+^1,\phi_+^2)}\right\}_0 +
\phi^2_+ \left\{\fr{W(\phi_+^1,\phi_-^1)}{W(\phi_+^1,\phi_+^2)} -  
\frac{8\xi\phi_-^1\phi_+^1 }{aW(\phi_+^1,\phi_+^2)}\right\}_0 &,
\rho>0,\\
\phi^1_- &, \rho<0,
\end{array}\right.
\adb\\
\tilde{\phi}^2 &=&\left\{\begin{array}{cl}
\phi^1_+\left\{-\fr{W(\phi_+^2,\phi_-^2)}{W(\phi_+^1,\phi_+^2)}  +
\frac{8\xi\phi_-^2\phi_+^2 }{aW(\phi_+^1,\phi_+^2)}\right\}_0  +
\phi^2_+ \left\{\fr{W(\phi_+^1,\phi_-^2)}{W(\phi_+^1,\phi_+^2)}  -
\frac{8\xi\phi_-^2\phi_+^1 }{aW(\phi_+^1,\phi_+^2)}\right\}_0&, \rho
>0,\\
\phi^2_- &, \rho<0,
\end{array}\right.
\adb\\
\hat{\phi}^1 &=&\left\{\begin{array}{cl}
\phi^1_+&,\rho>0,\\
\phi^1_-\left\{\fr{W(\phi_+^1,\phi_-^2)}{W(\phi_-^1,\phi_-^2)} -
\frac{8\xi\phi_-^2\phi_+^1 }{aW(\phi_-^1,\phi_-^2)}\right\}_0 +
\phi^2_-\left\{-\fr{W(\phi_+^1,\phi_-^1)}{W(\phi_-^1,\phi_-^2)} +
\frac{8\xi\phi_-^1\phi_+^1 }{aW(\phi_-^1,\phi_-^2)}\right\}_0&,\rho<0
\end{array}
\right.\adb\\
\hat{\phi}^2 &=&\left\{\begin{array}{cl}
\phi^2_+&,\rho>0,\\
\phi^1_-\left\{\fr{W(\phi_+^2,\phi_-^2)}{W(\phi_-^1,\phi_-^2)} -
\frac{8\xi\phi_-^2\phi_+^2 }{aW(\phi_-^1,\phi_-^2)}\right\}_0 +
\phi^2_-\left\{\fr{W(\phi_-^1,\phi_+^2)}{W(\phi_-^1,\phi_-^2)} +
\frac{8\xi\phi_+^2\phi_-^1 }{aW(\phi_-^1,\phi_-^2)}\right\}_0&,\rho<0.
\end{array}\right.
\end{eqnarray*}

In order to satisfy the boundary conditions for $\Psi$, namely $\lim_{\rho\to\infty}
\Psi_2 =0$, we have to take $\alpha^2_+=0$. For the second solution there
is no additional condition to leave only one constant. In the space-time without
wormhole we have the additional condition at the point $\rho = 0$ and therefore,
the functions must be finite. Here we have no origin, there is a bridge starting
from some distance of throat. For this reason we have to consider both
possibilities. Let us consider the specific solution for a second solution for
$\alpha^1_-=0$. It means that we consider the function which is symmetric to
$\Psi_2$, and tends to zero in the mirror spacetime for $\rho\to -\infty$. These
solutions have the following form:
\bs\label{psispec}
\begin{eqnarray}
\Psi_1 &=& \beta^1_- \tilde{\phi}^2, \adb \\
\Psi_2 &=&  \beta^2_+ \hat{\phi}^2.
\end{eqnarray}
\es
In what follows we will consider the symmetric profile of the throat $r(-\rho) =
r(\rho)$. Taking into account the above relations we obtain the radial
Green's function in the following form

\bs\label{g_lGen3}
1. $\rho>\rho'>0$
\be
\label{g^1Gen3}
g_l^{(1)}(\rho,\rho') =  -\fr 1{A_+}\phi^2_+(\rho)\phi^1_+(\rho') +
\left.\fr 1{A_+}\fr{W_+(\phi^1_+,\phi^2_+) + \frac{8\xi}a
\phi_+^2\phi_+^1}{W_+(\phi^2_+,\phi^2_+) + \frac{8\xi}a \phi_+^2\phi_+^2}
\right|_0 \phi^2_+(\rho')\phi^2_+(\rho)
\ee

2. $0<\rho<\rho'$
\be
g_l^{(2)}(\rho,\rho') = g_l^{(1)}(\rho',\rho)
\ee

3. $\rho < \rho'$ and $\rho'>0,\ \rho<0$
\be
g_l^{(3)}(\rho,\rho') = -\left.\fr
1{A_+}\fr{W(\phi^1_+,\phi^2_+)}{W_+(\phi^2_+,\phi^2_+) + \frac{8\xi}a
\phi_+^2\phi_+^2} \right|_0\phi^2_+(\rho')\phi^2_+(-\rho)
\ee

4. $\rho>\rho'$ and $\rho'<0,\ \rho>0$
\be
g_l^{(4)}(\rho,\rho')  = g_l^{(3)}(\rho',\rho')
\ee

5. $\rho' < \rho<0$
\be
g_l^{(5)}(\rho,\rho') = g_l^{(1)}(-\rho,-\rho')
\ee

6. $\rho < \rho'<0$
\be
g_l^{(6)}(\rho,\rho')  = g_l^{(5)}(\rho',\rho)
\ee
\es
In fact, we have to write out only $g_l^{(1)}$ and $g_l^{(3)}$ in manifest form.

The second kind of solutions has the following form 
\bs\label{psispec2}
\begin{eqnarray}
\Psi_1 &=& \alpha^1_- \tilde{\phi}^1, \adb \\
\Psi_2 &=&  \beta^2_+ \hat{\phi}^2.
\end{eqnarray}
\es
In this case the Green functions are given by

\bs\label{g_lSym2}
1. $\rho>\rho'>0$
\be\label{g^1}
g_l^{(1)}(\rho,\rho') =  -\fr 1{A_+}\phi^2_+(\rho)\phi^1_+(\rho') +
\fr 1{A_+} \left. \fr{W_+(\phi^1_+,\phi^1_+) + \frac{8\xi}a
\phi_+^1\phi_+^1}{W_+(\phi^1_+, \phi^2_+) + \frac{8\xi}a
\phi_+^1\phi_+^2}\right|_0 \phi^2_+(\rho')
\phi^2_+(\rho)
\ee

2. $0<\rho<\rho'$
\be
g_l^{(2)}(\rho,\rho') = g_l^{(1)}(\rho',\rho)
\ee

3. $\rho < \rho'$ and $\rho'>0,\ \rho<0$
\be
g_l^{(3)}(\rho,\rho') = -\left.\fr
1{A_+}\fr{W(\phi^1_+,\phi^2_+)}{W_+(\phi^1_+,\phi^2_+)+ \frac{8\xi}a
\phi_+^1\phi_+^2} \right|_0\phi^2_+(\rho')\phi^1_+(-\rho)
\ee

4. $\rho>\rho'$ and $\rho'<0,\ \rho>0$
\be
g_l^{(4)}(\rho,\rho')  = g_l^{(3)}(\rho',\rho')
\ee

5. $\rho' < \rho<0$
\be
g_l^{(5)}(\rho,\rho') = \fr 1{A_+}\phi^1_+(-\rho')\phi^2_+(-\rho) -
\left.\fr 1{A_+}\fr{W_+(\phi^2_+,\phi^2_+) - \frac{8\xi}a
\phi_+^2\phi_+^2}{W_+(\phi^1_+,\phi^2_+) + \frac{8\xi}a \phi_+^1\phi_+^2}
\right|_0 \phi^1_+(-\rho')\phi^1_+(-\rho)\label{g5}
\ee

6. $\rho < \rho'<0$
\be
g_l^{(6)}(\rho,\rho')  = g_l^{(5)}(\rho',\rho)
\ee
\es
For smooth background we have to set $\xi = 0$ in the above formulas except for
the differential equation for the radial Green function.

From relation (\ref{g5}) we observe that the second solution gives
a divergent Green function in the domain $\rho <0$ because the function
$g_l^{(5)}$ contains multiplication of functions which tends to infinity for
great $\rho$. For this reason we have to throw away this solution and consider them
as unphysical and thus, the Green function is uniquely defined.

\section{Self-energy and self-force}

Let us firstly call attention to the fact that for scalar particles it is possible to solve 
the problem concerning the determination of the self-force and self-energy in a closed form, only 
for an infinitely short throat and for the profile $\sqrt{\rho^2 + a^2}$, in massless case. For arbitrary
profile there is a problem to solve this problem for the zero mode $l=0$. Otherwise, for the electromagnetic 
case, this problem can be solved in a closed form for arbitrary profile.

\subsection{Profile $r=|\rho| + a$}
Let us first of all consider the simplest profile of the throat given
by $r=|\rho| + a$.
\subsubsection{Massless case}
In this case we may use the same solutions as the one in Ref.
\cite{Khusnutdinov:2007:Scpwst}. We have the equation
\begin{equation*}
 \phi'' + \frac{2r'}{r}\phi' - \frac{l(l+1)}{r^2}\phi = 0,
\end{equation*}
which has the following solutions
\begin{eqnarray*}
 \phi^1_\pm &=& (a\pm\rho)^l,\adb\\
 \phi^2_\pm &=& a^{2l+1}(a\pm\rho)^{-l-1}.
\end{eqnarray*}
Taking into account the above formulas we obtain the following expression for
the Green function
\begin{eqnarray*}
 (2l+1)g^{(1)}_l &=& \frac{r'^l}{r^{l+1}} - \frac{a^{2l+1}}{r^{l+1}r'^{l+1}}
\frac{1-8\xi}{2(l+1)-8\xi},\adb\\
(2l+1)g^{(3)}_l &=&\frac{a^{2l+1}}{r^{l+1}r'^{l+1}}
\frac{1-8\xi}{2(l+1)-8\xi}. 
\end{eqnarray*}
We use now the first expression above in the Eq. (\ref{GreenDefinition}) with
the coincidence limit for the angular variables, namely, $\gamma =0$, and thus we have
\begin{equation*}
 G(\rho;\rho') = \sum_{l=0}^\infty
\frac{2l+1}{4\pi} g_l(\rho,\rho'),
\end{equation*}
where $\rho'$ is the observation point and $\rho$ is the position of the
particle. It is very easy to make the summation over $l$. Doing this, we arrive at the
following results 
\begin{eqnarray}
 G(\rho,\rho') &=& \frac{1}{4\pi} \frac{1}{\rho -\rho'} - \frac{a(1-8\xi)}{8\pi
r r'} \Phi\left(\frac{a^2}{rr'},1,1-4\xi\right),\adb\\
U(\rho) &=& - \frac{ae^2(1-8\xi)}{4
r^2} \Phi\left(\frac{a^2}{r^2},1,1-4\xi\right)\label{UPhi}
\end{eqnarray}
for Green function and self-energy, respectively. The first term in the Green
function is the DeWitt-Schwinger expansion and we subtracted this term in accordance with the
discussion in the end of last section. The definition and properties
of function $\Phi$, 
\begin{equation}
 \Phi\left(\frac{a^2}{r^2},1,1-4\xi\right) = \sum_{n=0}^\infty (1-4\xi+n)^{-1}
\left(\frac ar\right)^{2n},
\end{equation}
may be found in Ref. \cite{Bateman:1953:Htf}.

The limiting cases $\rho\to \infty$  and $\rho\to 0$ gives us the following
results
\begin{eqnarray*}
\lim_{\rho \to \infty}U &=& -\frac{ae^2}{4\rho^2} \frac{1-8\xi}{1-4\xi},\adb\\
\lim_{\rho \to 0}U &=& \frac{e^2(1-8\xi)}{4a} (\ln \frac{2\rho}{a} + \gamma +
\Psi(1-4\xi)).
\end{eqnarray*}
For $\xi = 1/8$ we obtain zero as should be the case due to conformal
flatness of the $3D$ section of constant time. We observe that according with
the discussion above the divergence for $\xi =1/4$ appears. Furthermore we
observe appearance infinite number poles at points $\xi_n = (n+1)/4$. The
appearance of problems with the delta-like potential was
noted in literature 
\cite{Berezin:1961:rSesp,Mamaev:1982:Qeefdbppls,Bordag:1992:Veqftepcp,%
Albeverio:1988:Smqm}. In the limit of minimal coupling we
recover the result for the electromagnetic field given by
\begin{equation}
 U = \frac{e^2}{4a} \ln \left(1-\frac{a^2}{r^2}\right).
\end{equation}
Indeed, in the limit $\xi \to 0$ and for $m=0$ the equation for scalar field
(\ref{eq:phi}) coincides with the equation for component $A_0$ of vector field
and the formula for calculation the self-energy (\ref{Ureg}) coincides with
that for electromagnetic field (see Ref. \cite{Khusnutdinov:2007:Scpwst}).
Therefore, considering the $e$ as an electric charge we obtain that in this
limit we have to recover the electromagnetic case.

For the second kind solution given by (\ref{g5}) we obtain the following expression
\begin{eqnarray}
 (2l+1)g^{(5)}_l &=& -\frac{r'^l}{r^{l+1}} +
\frac{2l+1-8\xi}{1-8\xi}\frac{r^{l}r'^{l}}{a^{2l+1}}.
\end{eqnarray}
We observe that the series over $l$ is divergent because $r\geq a$ and
$r'\geq a$. Therefore, as noted above we have throw away this solution due to
the fact that it is unphysical. 

\subsubsection{Massive case}

In this case the radial equation turns to 
\begin{equation*}
 \phi'' + \frac{2r'}{r}\phi' - \left(\frac{l(l+1)}{r^2} + m^2\right)\phi =
0
\end{equation*}
and has two independent solutions, given in terms of the modified spherical
Bessel functions, $I_\nu$ and $K_\nu$, as
\begin{eqnarray*}
 \phi^1_+ &=& \sqrt{\frac{\pi}{2 x}}I_{\nu}(x),\adb\\
 \phi^2_+ &=& \sqrt{\frac{\pi}{2 x}}K_{\nu}(x),
\end{eqnarray*}
where $x= m(a+\rho)$ and $\nu = l+1/2$. Using these solutions we obtain
the Green function 
\bs
\begin{eqnarray}
 g_l^{(1)} &=& \frac{K_\nu(m r)I_\nu(m r')}{\sqrt{rr'}}  -
  \left.\frac{m a(I_\nu K_\nu' + I_\nu'K_\nu) +
(8\xi-1) I_\nu K_\nu}{2m a K_\nu K_\nu' + (8\xi-1)
K_\nu^2}\right|_{m a} \frac{K_\nu(m r)K_\nu(mr')}{\sqrt{rr'}} ,\adb\\
g_l^{(3)} &=& -\left.\frac{1}{2m a K_\nu K_\nu' + (8\xi-1)
K_\nu^2}\right|_{m a} \frac{K_\nu(m r)K_\nu(m r')}{\sqrt{rr'}} 
\end{eqnarray}
\es
where $r=|\rho|+a,\ r' = |\rho'| +a$. 

Using the addition theorem for Bessel function \cite{Bateman:1953:Htf-1} we
obtain the following formula ($r>r'$)
\begin{equation}
 \frac{1}{\sqrt{rr'}}\sum_{l=0}^\infty (2l+1) I_\nu (m r') K_\nu (m r) =
\frac{1}{r-r'}e^{-m (r-r')},
\end{equation}
and consequently the first term in $g_l^{(1)}$ represents the standard Yukawa
contribution. The second term can not be represented in a closed form and we have
to calculate it numerically. After renormalization we get the expression
($\rho >0$)
\bs
\begin{eqnarray}
 G_{ren}(\rho,\rho) &=& -\frac{1}{2\pi} \sum_{l=0}^\infty \nu
\left.\frac{m a(I_\nu K_\nu' + I_\nu'K_\nu) +
(8\xi-1) I_\nu K_\nu}{2m a K_\nu K_\nu' + (8\xi-1)
K_\nu^2}\right|_{m a} \frac{K_\nu^2(m r)}{r},\adb\\
U(\rho) &=& 2\pi e^2 G_{ren}(\rho,\rho).
\end{eqnarray}
\es

At the beginning we discussed the divergence at $\xi = 1/4$. Let us
consider in manifest form the first term ($l=0$) in the renormalized radial
Green function
\begin{equation*}
 g_{l,ren}^{(1)} = -
  \left.\frac{m a(I_\nu K_\nu' + I_\nu'K_\nu) +
(8\xi-1) I_\nu K_\nu}{2m a K_\nu K_\nu' + (8\xi-1)
K_\nu^2}\right|_{m a} \frac{K_\nu^2(m r)}{r}. 
\end{equation*}
It has the following forms
\begin{eqnarray}
 g_{0,ren}^{(1)} &=& \frac{-e^{-2 m (r-a)} (3-8 \xi )+e^{-2 m
r} (3-8
\xi +4 a m)}{4 m r^2 (1+a m-4 \xi )},\label{gm}\adb\\
g_{0,ren}^{(1)}|_{m\to 0} &=& -\frac{a}{2r^2} \frac{1-8\xi}{1-4\xi},
\end{eqnarray}
for massive and massless cases, respectively. We observe that there is no
singularity for massive case at point $\xi = 1/4$. Indeed, for $\xi = 1/4$ we
have
\begin{equation*}
 g_{0,ren}^{(1)} = \frac{-e^{-2 m (r-a)}  + e^{-2 m r} (1 +4 a
m)}{4 m^2 a r^2 }.
\end{equation*}
But this expression blows up for massless case because the expansion gives us 
\begin{equation*}
 g_{0,ren}^{(1)} = \frac{1}{2mr^2} + \ldots .
\end{equation*}
Therefore in the massive case the expression is no longer singular at point
$\xi = 1/4$. The singularity appears at point $1/4 + ma/4$. The next term with
$l=1$ will show a singularity at point $1/2 + (ma)^2/4(1+ma)$ and so on. In
general we observe that divergences appear as solutions of the following
equation
\begin{equation}
 2 x K'_\nu(x) + (8\xi -1)K_\nu(x) = 0.
\end{equation}
For massless case ($x=0$) there is general solution of this equation, $\xi_n =
(n+1)/4$, where $\nu = n+1/2$, and we obtain infinite number of poles. For
massive case we have infinite number of solutions, too, but it is
impossible obtain solution in closed form for arbitrary $n$. From above
consideration we may conclude that nonzero mass of field gives correction
for $\xi_n$ obtained for massless case.

\begin{figure}[ht]
\begin{center}
{\epsfxsize=8truecm\epsfbox{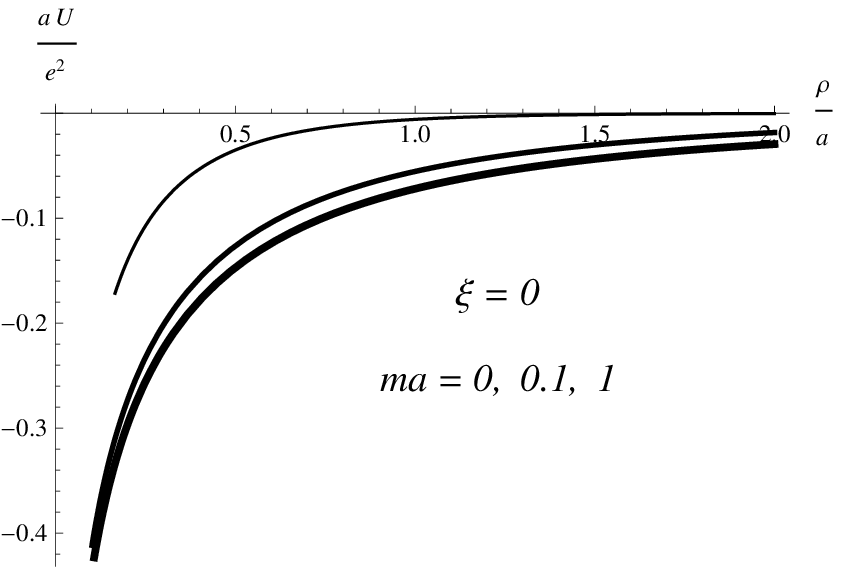}}%
\hspace*{2ex}{\epsfxsize=8truecm\epsfbox{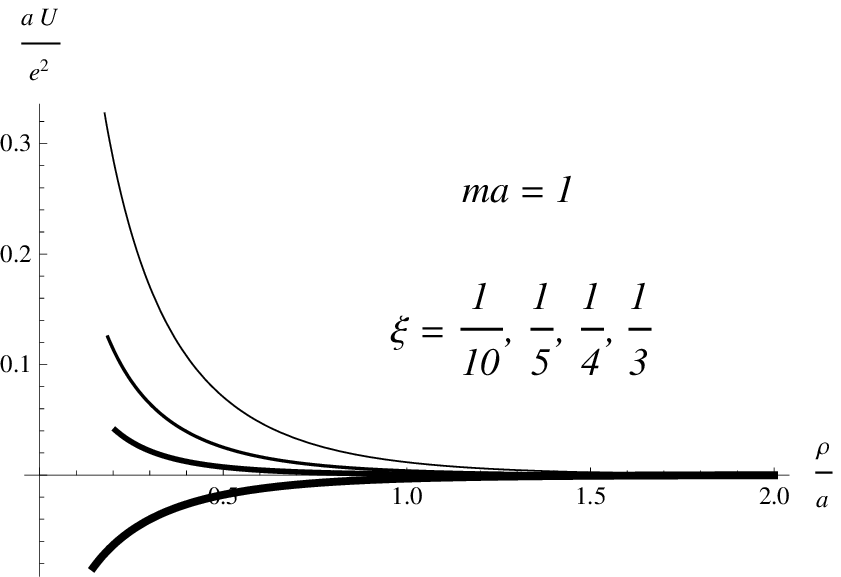}}
\end{center}\caption{The numerical simulation of the self-energy of a massive
scalar field for $\xi =0$ and for different parameters $ma=0$ (thick line), 
$ma=0.1$ (middle thickness) and $ma = 1$ (thin line). In the figure at right we
show the numerical simulation for $ma=1$ and for different parameters $\xi
=1/10$ (thick line) up to $\xi =1/3$ (thin line). For $\xi = 1/2$ it tends to
infinity as should be the case.}\label{fig:1}
\end{figure}

Let us consider the convergence of the series for the Green function, that is,
we have to consider the expressions for $\nu\to\infty$ and fixed $r$. With this
aim we use the uniform expansion for the Bessel function
\cite{Abramowitz:1970:eHMFFGMT} which is
valid for great index. We suppose that $mr/\nu$ and $ma/\nu$ are constants and
use the uniform expansion for functions $I_\nu (\nu z)$ and $K_\nu (\nu z)$. We
make expansion for $\nu g^{(1)}_{l,ren}$ over $\nu^{-1}\to 0$ and then make the
expansion of the obtained expression over $\nu\to\infty$ because each term of
the expansion depends on $\nu$ through $mr/\nu$ and $ma/\nu$. Doing this, we
obtain the following result
\begin{eqnarray}
 \nu g_l^{(1)} &=&
\frac{\left(\frac{a}{r}\right)^\nu}{r}\left\{-\frac{\zeta
}{4\nu}+\frac{\zeta}{8\nu^2} \left[2 m^2 \left(r^2-a^2\right)+\zeta \right] -
\frac{1}{16\nu^3}\left[4 a^2 m^2 \right.\right.\adb\nonumber\\
&+& \left.\left. 2 m^2 \left(a^4 m^2 - r^2-m^2 r^4 - a^2
\left(1+2 m^2 r^2\right)\right) \zeta + 2 m^2 \left(r^2 - a^2\right) \zeta
^2 +\zeta ^3\right]+\ldots\right\},
\end{eqnarray}
where $\zeta = 1-8\xi$. We observe that for $r>a$ the series is always
convergent. For $r=a$ the series is still convergent only for $\xi = 1/8$.
Therefore, the energy at the throat, $r=a$, is divergent for any case,
except for $\xi = 1/8$. The numerical simulations of the self-energy are shown
in Fig. \ref{fig:1}. 
We note that the massive field will produce a self-force which is
localized close to the throat. It falls down exponentially fast as $e^{-mr}$ far
from the throat. This behavior is in agreement with Linet result 
\cite{Linet:1986:wescs}.

\subsection{Profile $r = \sqrt{\rho^2 + a^2}$}

Because the mass of a field leads to suppression of the self-force by factor
$e^{-mr}$ we consider the massless case, $m=0$. The radial equation is written
as 
\begin{equation*}
 \phi'' + \frac{2\rho}{(\rho^2 + a^2)}\phi' -\left(\frac{l(l+1)}{\rho^2 +
a^2} - \frac{2\xi a^2}{(\rho^2 + a^2)^2}\right)\phi = 0,
\end{equation*}
which has two linearly independent solutions ($\mu =\sqrt{2\xi}$), namely,
$\phi^1_+$
and $\phi^2_+$, given by 
\bs\label{phi1}
\bn
\phi^{1}_+ &=& c_1^+P_l^{-\mu} (z),\adb\label{phi1_1} \\
\phi^{2}_+ &=& c_2^+Q_l^{\mu} (z),\label{phi1_2}
\en
\es
with Wronskian
\bn
W(\phi_+^{1},\phi_+^{2}) &=& ic_1^+c_2^+
\fr{a}{r^2}e^{i\pi\mu}\label{wronskian4}.
\en
Here $P_l^\mu$ and $Q_l^\mu$ are the Legendre functions of the first and
second kinds, and $z=i\rho/a$. Taking into account the above relations we obtain
the following expression for the Green function
\bn
G(\rho,\rho') &=& \frac{1}{4\pi a } \sum_{l=0}^\infty (2l+1)
\left\{ie^{-i\pi\mu} P_l^{-\mu} (z')Q_l^\mu (z) + \frac{e^{i\pi\mu}}{
\pi}Q_l^{-\mu} (z')Q_l^\mu (z)\right\}\adb\nonumber\\
&=& \frac{1}{4\pi a } \sum_{l=0}^\infty (2l+1)
\left\{  ie^{i\pi\mu}P_l^\mu (z')Q_l^{-\mu} (z) + \frac{e^{-i\pi\mu}}{\pi}
Q_l^{\mu} (z')Q_l^{-\mu} (z)\right\}.\label{green}
\en

As expected, for $\xi = 1/2$, that is for $\mu =\sqrt{2\xi} = 1$, we have a
divergence as $\Gamma (1-\mu)$ in the second term of the sum in (\ref{green})
for $l=0$. Indeed, for $l=0$ and $\mu =1$ we have 
\begin{eqnarray}
 P_0^{-1}(z) &=& \frac{\sqrt{z-1}}{\sqrt{z+1}},\\
 Q_0^1(z) &=& -\frac{1}{\sqrt{z-1}\sqrt{z+1}},
\end{eqnarray}
and the first term in the series reads
\begin{eqnarray}
 -iP_0^{-1}(z')Q_0^1(z) - \frac{1}{\pi} \frac{\Gamma (1-\mu)}{\Gamma (1+\mu)}
Q_0^1(z')Q_0^1(z).
\end{eqnarray}
Therefore, the Legendre's functions are finite but the gamma function is
divergent at $\mu=1$. We noted this critical value of $\xi$ from the
quantum mechanics point of view. Let us consider this moment in manifest form
the calculations starting from the beginning. For $l=0$ and $\mu =1$ we obtain
two solutions 
\begin{eqnarray}
 \phi^1_+ &=& \frac x{\sqrt{1+x^2}},\ \phi^2_+ = \frac 1{\sqrt{1+x^2}},\adb\\
 \phi^1_- &=& -\frac x{\sqrt{1+x^2}},\ \phi^2_- = \frac 1{\sqrt{1+x^2}},
\end{eqnarray}
where $x=\rho /a$. It is easy to see that the matching conditions give the
following solutions
\begin{eqnarray*}
\Psi &=&\left\{ \begin{array}{lc}
                  \frac{\alpha_+x}{\sqrt{1+x^2}} +
\frac{\beta_+}{\sqrt{1+x^2}}, & \rho >0 \\
                  \frac{\alpha_+x}{\sqrt{1+x^2}} +
\frac{\beta_+}{\sqrt{1+x^2}}, &
                  \rho <0
                \end{array} \right. .
\end{eqnarray*}
The first solution falls down for $\rho\to + \infty$ and the second solution
falls down for $\rho\to - \infty$. Therefore, we will get the following expressions
\begin{eqnarray*}
 \Psi_1 &=& \frac{\beta_+}{\sqrt{1+x^2}},\adb \\
 \Psi_2 &=& \frac{\beta_+}{\sqrt{1+x^2}},
\end{eqnarray*}
and the Wronskian is obviously zero. Because the solutions have the Wronskian in
the denominator we obtain a divergent result. 

Unfortunately, there is no closed expression for the above series for arbitrary
$\xi$. Nevertheless, for the specific case $\xi = 1/8$, we may obtain the Green
function and the self-force in manifest form. Indeed, in this case $\mu = 1/2$
and Legendre functions are expressed in terms of simple functions
\cite{Bateman:1953:Htf}, which helps us to obtain the following expression for
the Green function 
\bn
 G(\rho,\rho') &=& \frac{1}{4\pi a }
\frac{\sqrt{pp'}}{(p-p')(1+x^2)^{1/4}(1+x'^2)^{1/4}},
\en
where $p = x+\sqrt{x^2+1}$ and $x=\rho/a$. Then, we renormalize the Green
function and take the coincidence limit and obtain: 
\begin{equation}
 G^{ren}(\rho,\rho) = [G - \frac{1}{4\pi(\rho-\rho')}] = 0.
\end{equation}
Therefore, as expected, the self-force for $\xi = 1/8$ for this profile of
throat is zero. We will confirm this result by numerical calculation below. 

To obtain the expression for arbitrary $\xi$ we perform WKB analysis of the
radial equation
\be
\label{radial_}
\phi'' + \fr{2 r'}r \phi' - \left(\fr{l(l+1)}{r^2} + \xi R\right) \phi
= 0
\ee
and represent the solution of this equation in the form
\be
\phi = e^{S}.\label{e^S}
\ee
Thus, we obtain the following equation for $S$:
\be
S'+S'^2 + \fr{2 r'}r S' - \fr{\nu^2-1/4}{r^2} - \xi R
=0,\label{S}
\ee
where $\nu = l+1/2$. The next step is to expand $S$ in the following
power series in $\nu$ :
\be
S = \sum_{n=-1}^\infty \nu^{-n} S_n.\label{Sexp}
\ee

As noted in Ref. \cite{Khusnutdinov:2007:Scpwst} we have to
consider the term with $l=0$ separately. In this case the equation \Ref{radial_}
simplifies considerably ($\varphi$ stands here for the zero mode only):
\bd
\varphi'' + \fr{2 r'}r \varphi' - \xi R \varphi = 0.
\ed
The general solution of the above equation for our specific profile reads
\begin{equation}
 \varphi = C_1\cos(\mu\arctan\frac{\rho}{a}) +
C_2\sin(\mu\arctan\frac{\rho}{a}).
\end{equation}
Thus, we have the following solutions
\begin{eqnarray}
 \varphi^1_+ &=& \frac{2\tan\frac{\pi\mu}{2}}{\pi\mu}\left(
k_1\cos(\mu\arctan\frac{\rho}{a}) + k_2\sin(\mu\arctan\frac{\rho}{a})
\right),\adb\nonumber\\
\varphi^2_+ &=& \frac{\pi}2\left\{\cos(\mu\arctan\frac{\rho}{a}) -
\cot\frac{\pi\mu}2\sin(\mu\arctan\frac{\rho}{a})\right\}\adb\nonumber\\
&=& \frac{\pi}{2\sin \frac{\pi\mu}{2}} \sin \mu\left(\int_\rho^\infty
\frac{d\rho}{r^2}\right),\label{phi2secondmodel}
\end{eqnarray}
with Wronskian 
\begin{eqnarray}
W(\varphi_+^1,\varphi_+^2) &=& -\frac{a}{r^2}(k_1 + k_2
\tan\frac{\pi\mu}{2}).
\end{eqnarray}
Therefore we obtain expression
\begin{equation}
 g_0^{(1)} = \frac{\cos(2\mu \arctan \frac\rho a) - \cos (\pi\mu)}{2 a \mu
\sin\pi\mu}
\end{equation}
which does not depend on $k_1$ and $k_2$.

Now we consider terms with $l>0$. Substitution of \Ref{Sexp} into \Ref{S} yields
the set of expressions for the functions $S_n$. General solution of
the first four equations of this chain reads
\begin{eqnarray*}
 S'_{-1} &=& \pm \fr 1 r,\adb\\
 S'_0    &=& - \fr{r'}{2r} = -\fr 12 (\ln r)',\adb \\
 S'_1 &=& \pm\fr{\zeta}{4} \left[ r'' + \fr{r'^2}{2r} -
\fr{1}{2r}\right],\adb \\
 S'_2 &=& -\fr{\zeta}8 \left[r r^{(3)} + 2 r' r''\right]= -
\fr{\zeta}8 (r r'' + \fr 1{2} r'^2)^\cdot, \adb\\
 S'_3 &=& \frac{\zeta}{16} \left[r^{(4)} r^2+2 r''^2 r+4 r' r^{(3)} r+2
   r'^2 r''\right]-\frac{\zeta ^2 }{128 r}\left[r'^2+2 r r''-1\right]^2,\adb\\
 S'_4 &=& \frac{\zeta^2}{32} \left[r'^2+2 r r''-1\right]
   \left[2 r' r''+r r^{(3)}\right]\adb\nonumber\\
&+& \frac{\zeta}{32} \left[-2 r''
r'^3-10 r r^{(3)} r'^2 - r \left(8
   r''^2 + 7 r r^{(4)}\right) r'-r^2 \left(8
   r'' r^{(3)}+r r^{(5)}\right)\right]\adb\nonumber\\
&=& \frac{\zeta ^2}{128}  \left[r'^4 + 2\left(2
   r r''-1\right) r'^2+4 r r'' \left(r
   r''-1\right)\right]^\cdot \adb\nonumber\\
&-&\frac{\zeta r }{32}   \left[2
   r'' r'^2+4 r r^{(3)} r'+r \left(2
   r''^2+r r^{(4)}\right)\right]^\cdot .
\end{eqnarray*}
We observe that (i) all $S'_k$ with $k\geq 1$ are proportional to $\zeta =
1-8\xi$, and (ii) all $S'_{2k}$ are full derivative as in the case of $\xi=0$. 

\begin{figure}[ht]
\begin{center}
{\epsfxsize=8truecm\epsfbox{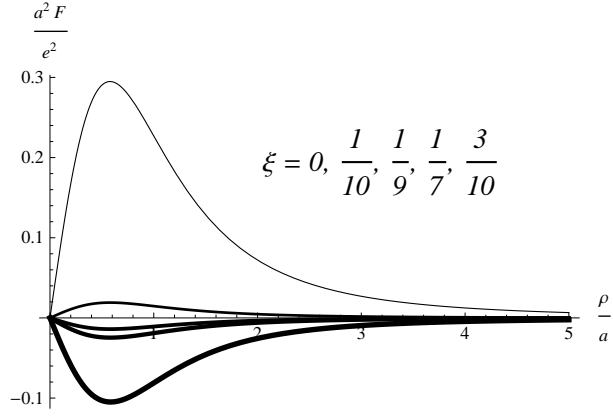}}
\end{center}\caption{The numerical simulation of the self-force on a massless
scalar field for profile $r=\sqrt{\rho^2 + a^2}$ for different parameters from
$\xi=0$ (thick line) up to $\xi = \frac{3}{10}$ (thin line). For $\xi = \frac
18$ it is zero and for $\xi = \frac 12$ it tends to infinity.}\label{fig:2}
\end{figure}

Now we are in position to calculate the Green function. First of all we
calculate the Wronskian and found
$A_+$:
\begin{equation}
 A_+ = W(\phi^1_+,\phi^2_+) r^2 = e^{S^{+2}+S^{+1}}(S'^{+2}-S'^{+1})
r^2. 
\end{equation}
By using the formulas above  we obtain
\begin{eqnarray}
 g^{(1)}_l(\rho,\rho') &=& \frac{e^{-\nu\int_{\rho'}^\rho
\frac{d\rho}{r}}}{2\nu\sqrt{r(\rho)r(\rho')}}  \frac{e^{-\sum_{n=1}^\infty
\nu^{-n}(S^{+1}_n(\rho) - S^{+1}_n(\rho'))}}{\sum_{n=0}^\infty \nu^{-2n}r(\rho
S'^{+1}_{2n-1}(\rho))}. 
\end{eqnarray}
Then, we change summation over $n$ and $l$ and get the following result  
\bnn
\sum_{l=0}^\infty \fr {2\nu}{4\pi}g_l(\rho,\rho') 
&=&\fr 1{4\pi} \fr{1}{\sqrt{r(\rho) r(\rho')}}\sum_{k=0}^\infty
f_k(b) j_k (\rho,\rho')+\fr 1{4\pi} g_0(\rho,\rho'),
\enn
where
\bnn
\fr{f_0(b)}{\sqrt{r(\rho)r(\rho')}} &=& \fr 1{\rho-\rho'} - \fr 1r +
O(\rho-\rho'),\adb \\
\fr{f_1(b)}{\sqrt{r(\rho)r(\rho')}} &=& -\fr 1 r \ln \fr{\rho
-\rho'}{4r} - \fr 2r + O(\rho-\rho'),\adb \\
\fr{f_k(b)}{\sqrt{r(\rho)r(\rho')}} &=& \fr 1r \zeta_H(k,\fr 32) +
O(\rho-\rho').
\enn
The functions $j_k$ may be found by using simple code in package
"Mathematica". There is no general form of these coefficients for arbitrary
index but for numerical calculations we need only for some first ones (see
Ref. \cite{Khusnutdinov:2007:Scpwst}). The first four coefficients have the
following form in coincidence limit
\bnn
j_0 (\rho,\rho') &=& 1,\adb \\
j_1 (\rho,\rho') &=& -\zeta\int_{\rho'}^\rho \fr{-1+ r'^2 + 2rr''}{8r} d\rho = -
\zeta\fr{-1+ r'^2 + 2r r''}{8r}(\rho-\rho') +
O((\rho-\rho')^2),\adb\\
j_2 (\rho,\rho) &=& - \zeta\fr{-1+ r'^2 + 2r r''}{8},\adb\\
j_4 (\rho,\rho) &=& \frac{3\zeta ^2}{128}  \left(r'^2+2 r
   r''-1\right)^2 - \frac{r \zeta}{16}   \left(2 r''
   r'^2+4 r r^{(3)} r'+r \left(2
   r''^2+r r^{(4)}\right)\right).
\enn
Therefore we obtain
\bnn
\sum_{l=0}^\infty \fr {2\nu}{4\pi}g_l(\rho,\rho') &=& \fr 1{4\pi}
\left[\fr 1{\rho-\rho'} - \fr 1r + \fr 1r \sum_{k=1}^\infty
\zeta_H(2k,\fr 32) j_{2k} (\rho,\rho) + \frac{\cos(2\mu \arctan \frac\rho a) -
\cos (\pi\mu)}{2 a \mu
\sin\pi\mu}\right],
\enn
where each term $j_{2k}$ is proportional to $\zeta = 1-8\xi$. Here
$\zeta_H (s,p)$ is the Hurwitz zeta function (see, for example
\cite{Bateman:1953:Htf}). After
regularization we arrive at the following formula for the self-energy ($\mu =
\sqrt{2\xi}$) 
\be
\label{U_renorm'd}
U(\rho) = \fr{e^2}{2} \left[- \fr 1r + \fr 1r \sum_{k=1}^\infty
\zeta_H(2k,\fr 32) j_{2k} (\rho,\rho)+ \frac{\cos(2\mu \arctan \frac\rho a) -
\cos (\pi\mu)}{2 a \mu
\sin\pi\mu}\right].
\ee
As expected it is zero for $\xi = 1/8$ and it is
divergent for $\xi = 1/2$. Far from the throat we obtain
\begin{equation}
 U \approx -\frac{e^2}{2\rho^2} \frac{a\mu}{\tan \pi\mu}.\label{Ufar}
\end{equation}
By numerical analysis it is enough to take into account only two terms of the
series above. In fact we use half of sum the first two terms. The numerical
simulations are reproduced in Fig. \ref{fig:2}.

For arbitrary profile of the wormhole we have the following formula
\be
U(\rho) = \fr{e^2}{2} \left[- \fr 1r + \fr 1r \sum_{k=1}^\infty
\zeta_H(2k,\fr 32) j_{2k} (\rho,\rho)+ g_0^{(1)}(\rho)\right],\label{Ugeneral}
\ee
with the same $j_{2k}$ as above and  
\begin{eqnarray}
 g_0^{(1)}(\rho,\rho) &=&-\fr 1{A_+}\varphi^2_+(\rho)\varphi^1_+(\rho) + \fr
1{2A_+} \left(
\fr{\varphi^1_+}{\varphi^2_+} +
\fr{\varphi'^1_+}{\varphi'^2_+}\right)_0
\varphi^2_+(\rho)\varphi^2_+(\rho),\label{ZeroModeGeneral}
\end{eqnarray}
where $A_+= W_+(\varphi^1_+,\varphi^2_+)r^2(\rho)$. The functions
$\varphi^{1,2}_+$ are the solutions of the equation
\be
\varphi'' + \fr{2 r'}r \varphi'  - \xi R \varphi = 0.
\ee
Unfortunately, differently from the electromagnetic field case, there is no
general solution of this equation for arbitrary $\xi$ and $r$. For $\xi = 1/8$
it is easy to find a general solution of this equation by using the conformal
flatness of the equation. They read
\begin{equation}
 \varphi^1 = \frac{1}{\sqrt{r}} e^{\int^\rho \frac{dy}{r(y)}},\ \varphi^2 =
\frac{1}{\sqrt{r}} e^{-\int^\rho \frac{dy}{r(y)}}, 
\end{equation}
with Wronskian $W(\varphi^1,\varphi^2) = -1/r^2$. For $\xi \not = 1/8$ we may
only make conclusion about  behavior of the self-force far from the wormhole's
throat. Indeed, changing function by the relation $\varphi = w/r$ we obtain 
\be
w''  - \left( \fr{r''}r + \xi R \right) w = 0.
\ee
Therefore, for great distance we have simple equation $w'' = 0$, with two
solutions 
\begin{equation}
 w_1 = c_2 \rho,\ w_2 = c_1. 
\end{equation}
The Wronskian corresponding solutions is 
\begin{equation}
 W(\varphi_1,\varphi_2) = - \frac{c_1c_2}{r^2}.
\end{equation}
It is not difficult to show that the solutions with next corrections are
\begin{eqnarray}
 \varphi_1 &=& c_1 \left(1 +
\frac{h_1}{\rho} + O(\rho^{-2})\right)\adb\nonumber, \\
\varphi_2 &=& \frac{c_2}\rho \left(1 + O(\rho^{-2})\right),
\end{eqnarray}
and therefore we obtain for great $\rho$ 
\begin{equation}
 U \approx -\frac{e^2}{2\rho^2} \left(
 -h_1 + \frac{c_2}{2c_1}\left(\frac{\varphi_1}{\varphi_2} +
\frac{\varphi'_1}{\varphi'_2}\right)_0\right)\label{uinf}.
\end{equation}
But we can not make any conclusion about sign of these expression. 

\section{Discussion and Conclusion}

In this paper we considered in details the self-interaction on a scalar particle
at rest in the wormhole space-time with non-minimal coupling with curvature.
The main peculiarities of the self-force on a scalar field are (i) mass of
field and (ii) nonminimal coupling $\xi$. We consider a particle at rest and for
this reason all equations become effectively three dimensional, because they
touch only spacial part of wormhole space-time which is conformally flat. For
this reason for $\xi = 1/8$ and massless field we expect that the self-force is
zero 
\cite{Hobbs:1968:Rdcfu}. Our calculations confirm this result, the self-force
zero indeed in all considered above special examples.  For $\xi < 1/8$ the
scalar particle is attracted to the wormhole and for $\xi>1/8$ the particle is
repelled by the wormhole. For $\xi = 1/8$ the self-force is zero. In
the space-time of a black hole \cite{Zelnikov:1982:Egpecp} one has different
behavior of the self-energy, in which case it is proportional to $\xi$ and the
self-force is zero for minimal coupling $\xi= 0$.

The self-force for scalar massless particles reveals peculiarity for specific
values of the nonminimal coupling $\xi_c$. The energy has a simple pole,
$(\xi -\xi_c)^{-1}$, at this point. Because the self-energy is defined in terms 
of the three dimensional Green function we may analyse this pole by using
analogy with the scattering theory \cite{Baz:1969:Srdnqmrirvnkm}. The
combination $V = \xi R + r''/r$ plays the role of a potential for the wave
function in non-relativistic quantum mechanics and all information about
boundary or scattering states is encoded in the Green function. This
potential tends to a constant at the wormhole's throat, $\rho=0$, and it falls
down to zero far from the throat. The critical value depends on the shape of
the throat. If the space-time far from the throat differs from Minkowski
spacetime as $b_n\rho^{-n}$, then the critical value $\xi_c = (n+1)/4n$ and the
potential, $V = n b_n (\xi_c -\xi)\rho^{-n-3}$  changes its sign in this point.
If the spacetime goes to Minkowski spacetime exponentially fast as $c_n\rho^n
e^{-\rho/\tau}$ then the critical value $\xi_c = 1/4$ and the potential, $V =
4c_n(\xi_c -\xi)\tau^{-2} \rho^{n-1} e^{-\rho/\tau}$, also changes its sign at
this point. This kind of wormholes possesses a dimensional parameter, $\tau$,
which may be regarded as length of the throat. The profile $r=|\rho| + a$ gives
singular, delta like potential consentrated at throat which has no longer
possess the length of the throat. This kind of profile belongs to the 
second type of throat due to localization potential close to the throat. We
obtain the same critical value $1/4$ in this case by manifest calculations
(see Eq. (\ref{UPhi})). 

For wormhole with profile $r(\rho ) = \sqrt{\rho^2 + a^2}$ the Green function
for zero mode, $l=0$, is expressed in terms of two solutions of radial equation
\begin{equation*}
 \Phi'' - \frac{a^2 (1-\mu^2)}{(\rho^2 + a^2)^2} \Phi = 0,
\end{equation*}
with $\mu^2=2\xi$. This equation describes the particle in the potential $V = -
{a^2 (1-\mu^2)}/{(\rho^2 + a^2)^2}$ in one dimension. It is well-known
\cite{Baz:1969:Srdnqmrirvnkm} that the Green function has the poles for energy
of boundary states. The point $\mu =1 (\xi = 1/2)$ is critical because the
potential changes its sign and boundary states appears or disappears. 

The mass of the field gives additional factor $e^{-mr}$ and leads to
localization of the self-force close to the wormhole's throat inside a sphere
with the Compton wavelength radius $m^{-1}$ (see Fig. \ref{fig:1}). The
self-force reveals singularity too but the point depends on the mass of the
field. For example, for a simple profile of the throat, $r=|\rho| + a$, the
singularity appears at the point $\xi_c = 1/4 + ma/4$ (\ref{gm}), where $a$ is
radius of the throat. 

We developed a procedure and found general expression (\ref{Ugeneral}) for
the self-force for general profile of the throat. But differently from the 
electromagnetic field case there is no general solution for zero mode
$g_0^{(1)}$ (\ref{ZeroModeGeneral}) in terms of profile function $r$. We would
like to note that this relation contains two independent solutions of
homogeneous radial equation for zero mode at observation point as well as at the
origin, for $\rho =0$. Because the fall down function $\varphi^2$ and its
derivative appears at the denominator we have to use irregular solution for this
function in terms of the scattering theory (see, for example
\cite{Newton:1966:STWP}).  We observe that zero mode gives main
contribution to self-force and it depends on the global structure of
space-time which is in agreement with consideration the self-force in black
hole background \cite{Linet:1976:EmSm,Vilenkin:1979:Scpgf}. For example, for
profile function $r = \sqrt{\rho^2 + a^2}$ we observe that zero mode solutions
given by Eq. (\ref{phi2secondmodel}) contains the integral over profile function
and therefore is defined by global structure of the space-time. The numerical
evaluations (see Fig. \ref{fig:2}) for this profile show that the self force as
expected changes its sign at the point $\xi = 1/8$: for $\xi < 1/8$ it is
attractive and for $\xi < 1/8$ it is repulsive. For $\xi \to1/2$ the self-force
reveals singularity as a simple pole $(\xi - 1/2)^{-1}$ and therefore tends to
infinity. The self-force has extrema for $\rho \approx a$ and it is zero at
origin. Far from the string the self-energy is given by Eq. (\ref{Ufar}).
Therefore we may say the same conclusion, the scalar particle will be
concentrated at the throat for $\xi < 1/8$ as for electromagnetic field case 
\cite{Khusnutdinov:2007:Scpwst}.

\section{Acknowledgments}
We would like to thank D. Chistyakov ans A. Popov for many discussions and
helpful comments on the paper. N.K. is grateful also to Departamento de
F\'isica, Universidade Federal da Para\'iba, Brazil, where this work was done,
for their hospitality. This work was supported in part by the Russian Foundation
for Basic Research, Grant No. 08-02-00325-a and in part by FAPESQ-PB(PRONEX)/CNPq, FAPES(PRONEX)/CNPq and 
by Conselho Nacional de Desenvolvimento Cientifico e Tecnol\'ogico CNPq, Brazil.

\end{document}